\documentstyle[prb,aps,twocolumn,epsf]{revtex}

\newcommand{\gapprox}{\mbox{\raisebox{-4pt}{$\,\buildrel>\over\sim\,$}}}
\begin{document}
\draft
\title{On the effect of irrelevant boundary scaling operators} 
\author{R.~Egger$^1$,  A.~Komnik$^1$, and H.~Saleur$^2$}
\address{
${}^1$Fakult\"at f\"ur Physik, Albert-Ludwigs-Universit\"at,
 D-79104 Freiburg, Germany\\
${}^2$Department of Physics,
University of Southern California, Los Angeles, CA 90089-0484}
\date{Date: \today}
\maketitle
\begin{abstract}
We investigate consequences of adding irrelevant (or less relevant) boundary
operators to a $(1+1)$-dimensional field theory,
using the Ising and the boundary sine-Gordon model
as examples. In the integrable case, irrelevant perturbations 
are shown to multiply reflection matrices by CDD factors: 
the low-energy behavior is not changed, while various high-energy behaviors 
are possible, including ``roaming'' RG trajectories. 
In the non-integrable case, a Monte Carlo study shows that the 
IR behavior is again generically 
unchanged, provided scaling variables are appropriately renormalized.
\end{abstract}
\pacs{}

\narrowtext

In this paper we study the effect of adding irrelevant or less relevant
scaling operators to a given theory, focusing on boundary perturbations
in a $(1+1)$-dimensional field theory.  This is a problem of important 
physical interest. Consider for instance tunneling in the fractional 
quantum Hall effect, which has been intensely studied recently
 in the context of Luttinger liquids, shot noise and 
fractional charge measurements. 
For filling factor $\nu=1/3$,
 there is a single relevant tunneling operator, but there are 
two of them  for  $\nu=1/5$. Most analytical calculations,
in particular the exact solutions, do hold only when the second,
 less relevant operator has been scaled away. In practice, however, 
this operator will always be there, and it is important to be able to 
evaluate its role.

The standard 
expectation is simply that one can neglect all less relevant and 
irrelevant  operators
when computing the properties close to the IR fixed point.\cite{Cardy}
This is, however, entirely based on weak coupling expansions, which 
can be quite misleading.\cite{ALT} For instance, 
the  added perturbations could become relevant at
strong coupling, entirely changing the physics in a non-perturbative way. 

Our study  supports the standard expectation
in a variety of cases related with the tunneling problem.
 On the conceptual side, one of our main results is
 that, in the integrable case,
 additional irrelevant perturbations contribute so-called CDD 
(Castillejo-Dalitz-Dyson)  factors to 
the boundary reflection matrices $R$. This is not so surprising: 
as discussed in Ref.~\onlinecite{GS}, $R$
is fully determined up to such CDD 
factors by the boundary analogues of the
Yang-Baxter equation,  unitarity and crossing symmetry,
and our results thus provide a natural explanation for the
CDD ambiguity.  Nevertheless, unexpected features are found close to the 
UV fixed point which can be altered by the added perturbations 
and is approached along a ``roaming''  
renormalization group (RG) trajectory that oscillates between
different boundary conditions. The relation 
between irrelevant perturbations and CDD factors has been 
discussed independently for the bulk case in Ref.~\onlinecite{MS}. 

To start, we consider the simplest possible situation, which is 
the scaling limit of the 
Ising model with zero external field but with a 
boundary magnetic field $h$. This integrable model is described by 
a free Majorana fermion field theory.\cite{GS} Since 
the  boundary spin operator coincides with the free 
fermion operator (in sharp contrast to the bulk case), 
 the problem can  be handled  by solving
the boundary equation of motion, which in turn determines 
 the boundary reflection matrices. 
It is then easy to include additional irrelevant 
operators: all  perturbations that are quadratic in the fermions will still
 lead to a solvable model.  

We illustrate this  for a boundary perturbation 
made up of two terms,  the spin operator
and the stress energy tensor $T=\pi :\psi\partial\psi:$, 
where $\partial\equiv \partial_z$ and $z=x+iy$.
The Euclidean action is
\begin{eqnarray} \label{isingact}
{\cal A} &=&\int_{-\infty}^\infty dy\int_{-\infty}^0 dx
\left(\psi\bar{\partial}\psi
-\bar{\psi}\partial\bar{\psi}+m\psi\bar{\psi}
\right)\\ \nonumber 
&+&{1\over 2}\int_{-\infty}^\infty dy\left(\psi\bar{\psi}(x=0,y)+{1\over i}a
\dot{a}(y)
\right)\\ &+& \nonumber
{h\over 2i}\int_{-\infty}^\infty dy \left(\psi+\bar{\psi}\right)(x=0,y)a(y)
\\ \nonumber &+&
{\lambda\over i}\int_{-\infty}^\infty dy\left(\psi\partial_y\psi+\bar{\psi}
\partial_y\bar{\psi}\right)(x=0,y) \;,
\end{eqnarray}
where $a$ is a fermionic boundary degree of freedom. The
case $\lambda=0$ was studied in Ref.~\onlinecite{GS}.
When $h=\lambda=0$, the model is at the fixed point 
corresponding to free boundary conditions, or in the fermion language, 
$\psi=\bar{\psi}$. It 
follows that on the boundary, $T=-i\pi:\psi\partial_y
\psi:$, and the last term in Eq.~(\ref{isingact})
reads $(\lambda/ \pi) \int dy\, (T+\overline{T})$.  

Let us now determine $R$.
The equations of motion at the boundary are
\begin{equation} \label{eqsofmot} 
{h^2\over 2i}\left(\psi+\bar{\psi}\right)
={d\over dy}\left(\psi-\bar{\psi}\right)-
2\lambda{d^2\over dxdy}\left(\psi-\bar{\psi}\right)\;.
\end{equation}
The bulk theory in Eq.~(\ref{isingact}) contains one
type of particle, namely the free fermion $A$ with mass $m$.
The fermion fields can be expressed in terms of the
particle-creation operator $A^\dagger(\theta)$,\cite{GS}
\begin{eqnarray}\label{modes}
\psi(x,y)&=&\int_{-\infty}^\infty d\theta\Bigl [
\omega e^{\theta/2} A(\theta) e^{imx\sinh\theta -my\cosh\theta}\\
\nonumber &+&
\bar{\omega}e^{\theta/2} A^\dagger(\theta) 
e^{-imx\sinh\theta+my\cosh\theta}\Bigr] \;, \\ \nonumber
\bar{\psi}(x,y)&=&\int_{-\infty}^\infty d\theta\Bigl[
\bar{\omega} e^{-\theta/2} A(\theta) e^{imx\sinh\theta -my\cosh\theta}
\\ &+& \nonumber
\omega e^{-\theta/2} A^\dagger(\theta) 
e^{-imx\sinh\theta+my\cosh\theta}\Bigr] \;,
\end{eqnarray}
where $\omega=\exp(i\pi/4), \bar{\omega}=\exp(-i\pi/4)$.
Inserting this expansion into Eq.~(\ref{eqsofmot}) and using the
defining relation $A^\dagger(\theta)  
=R(\theta) A^\dagger(-\theta) $,
it is straightforward to obtain the reflection matrix.
For simplicity, we focus on the massless limit,
$m\to 0$. In that limit, we boost the rapidities
$\theta$ such that the energy of the
particles $m\cosh\theta\to \exp(\beta)$, while the momentum 
$m\sinh\theta\to \pm \exp(\beta)$ for right- or left-movers,
respectively.\cite{foot0}  
Physical rapidities are characterized 
by $\exp(\beta)\approx T$, where $T$ denotes the temperature.

The reflection matrix for rapidity $\beta$ then reads
\[
R(\beta)=-i \, \frac{h^2+2ie^\beta+4\lambda e^{2\beta}}
{h^2-2ie^\beta+4\lambda e^{2\beta}}\;.
\]
This can be recast into the form
\begin{equation}\label{equivform}
R(\beta)=-i\,\tanh\left({\beta-\beta_B^1\over 2}-{i\pi\over 4}\right)
\coth\left({\beta-\beta_B^2\over 2}-{i\pi\over 4}\right)\;,
\end{equation}
where (the $+$ sign corresponds to $\beta_B^2$)
\begin{equation}\label{parametr}
\exp(\beta_B^{1,2})={\sqrt{1+4\lambda h^2}\pm 1\over 4\lambda}\;.
\end{equation}
Sending the additional irrelevant coupling to zero, $\lambda\to 0$,
we obtain $\exp(\beta_B^1)\approx h^2/2$, while 
$\exp(\beta_B^2)\approx 1/2\lambda$ diverges. 
Hence only the first term of the $R$ matrix (\ref{equivform})
matters, while the second factor saturates at minus one:
the boundary magnetic field $h$
induces a flow from free to fixed boundary conditions,\cite{GS} 
or, in the fermion language, 
from $\psi=\bar{\psi}$ to $\psi=-\bar{\psi}$.
Close to the UV fixed point ($\beta\to \infty$), 
the reflection matrix is $R_{free}=i$.   Close to the IR
fixed point ($\beta\to- \infty$), we instead get $R_{fixed}=-i$.

Let us now discuss the meaning of Eq.~(\ref{equivform}).
In the limit where  both bare couplings are small, 
we have $\exp(\beta_B^1)\ll \exp(\beta_B^2)$.
The vicinity of the free fixed point now corresponds to 
temperatures $\exp(\beta_B^1)\ll T\ll \exp(\beta_B^2)$.
In that regime, the $\coth$-factor in Eq.~(\ref{equivform})
 is close to minus one,
and Eq.~(\ref{isingact}) describes an Ising model with
free boundary conditions perturbed 
by a relevant ($\propto h$) and an irrelevant 
($\propto \lambda$) term. There are now two possibilities.
If we look at higher energies, i.e., run the RG backwards, 
the contribution of the irrelevant term is seen to {\sl grow} while the 
the relevant term disappears.  
Equation (\ref{equivform}) for the $R$ matrix shows that in this limit 
 the model flows to {\sl fixed}\, boundary conditions, $R\to -i$.
Therefore the UV fixed point is not characterized by free boundary
conditions anymore, as one might have expected.
If instead we look at lower energies, the irrelevant term is scaled away 
while the relevant term grows. 
From Eq.~(\ref{equivform}) we see that the model
again flows to fixed boundary conditions, $R\to -i$. 
The result of running the RG forward (to lower energies) is very natural
and confirms the general idea that irrelevant couplings should scale away.
The result of running the RG backwards (to higher energies) is less
natural, and its simplicity is probably connected to
the fact that we are dealing with a free theory. 
In more complex cases, we do not expect that such a backward RG trajectory
ever converges.

The structure of Eq.~(\ref{equivform})
generalizes to more complex mixtures of $N$  
perturbations that preserve integrability. Up to a factor $-i$, the $R$ 
matrix should thereby result as a product of $N$ (minus) hyperbolic tangents
with associated characteristic ``Kondo'' temperatures,
$\exp(\beta_B^1)\ll \exp(\beta_B^2) \ll \ldots \ll \exp(\beta_B^N)$.
If one considers some physical property as a function of  temperature, 
it will exhibit ``non-universal'' properties for $T\gapprox\exp(\beta_B^2)$.
For $T\ll \exp(\beta_B^2)$, however,
all the hyperbolic tangents but the first one are saturated, 
and then the reflection matrix becomes $
R\simeq i \tanh\left({\beta-\beta_B^1\over 2}-{i\pi\over 4}\right)$,
identical to the case of a single relevant perturbation. Note
that in these calculations, no cut-off has ever appeared. This 
is because of the implicit dimensional 
(normal-order) regularization used in integrable systems. 
In practice,
the theory must often be regularized using a UV cut-off, whence
a different  renormalization of $\beta_B^1$ by the irrelevant couplings 
will occur. 

Running the RG backwards from physical temperatures in the
range $\exp(\beta_B^1)\ll T \ll \exp(\beta_B^2)$
gives intriguing results. 
If $N$ is even, the model starts at free boundary conditions
and ends up at fixed ones.  The UV fixed point is approached
by ``roaming'', where the RG trajectory comes again
$N/2-1$ times close to free boundary conditions. 
For instance, if $N=4$, the backward RG trajectory 
reads: fixed $\leftarrow$ free $\leftarrow$ fixed $\leftarrow$ free. 
If $N$ is odd, the model ends up at free
boundary conditions, and comes close to fixed boundary conditions
 $(N-1)/2$ times. For instance, if $N=3$,
the backward trajectory reads:
free $\leftarrow$ fixed $\leftarrow$ free. Of course,
in a given realistic situation, we expect the 
equivalent of $N$ to be infinite,
so the backwards trajectory will keep roaming 
forever as energies get higher and higher. 
In a cut-off regularized theory, such a behavior will only
be meaningful up to energy scales given by the UV cut-off.
In contrast, the forward trajectory always describes the 
same flow from free to fixed boundary conditions.   

The above considerations for the Ising model  
can now be applied to the boundary sine-Gordon (BSG) model,
which in turn is related to the problem of a quantum impurity
in a Luttinger liquid,\cite{KF}
\begin{eqnarray}\label{bdrsg}
{\cal A}&=&{1\over 2}\int_{-\infty}^0dx\int_{-\infty}^\infty
dy \left[(\partial_x\Phi)^2+(\partial_y\Phi)^2\right] \\ \nonumber &+&
h\int_{-\infty}^\infty dy \cos[\sqrt{2\pi g}\Phi(x=0,y)] \;.
\end{eqnarray}
Here $\Phi(x,y)$ is a folded bosonic field.\cite{HS}
Remarkably, the $g=1/2$
BSG model can be mapped on a pair of Ising models.\cite{AKL,LLS} 
The first of these Ising models
has fixed boundary conditions $(h=\infty$), while the other sees an 
effective boundary magnetic field proportional to 
the coupling $h$. 
This mapping can be made explicit at the level of 
reflection matrices.\cite{LLS}
At $g=1/2$, the BSG model has kinks and antikinks 
with a bulk scattering matrix $S=-1$.  At the boundary,
they elastically scatter one by one without particle production.
The reflection amplitudes for kink-kink scattering, $P=R_{++}$,
 and kink-antikink scattering, $Q=R_{+-}$, are\cite{LLS}
\begin{equation}\label{corresp}
P+Q=-i\;,
\quad P-Q=i\,\tanh\left({\beta-\beta_B\over 2}-{i\pi\over 4}\right)\;, 
\end{equation}
where $\exp(\beta_B)\propto h^2$. The mapping onto the Ising models 
is obtained by taking  symmetric and antisymmetric 
linear combinations of kink and antikink, which then 
correspond to the fermions $\psi$ and $\xi$ of the two Ising models.
Clearly, the reflection matrices (\ref{corresp}) are consistent 
with the  Ising $R$ matrix (\ref{equivform}).

Suppose now that we add a general combination of the two
irrelevant operators $\cos [2\sqrt{2\pi g}\Phi(x=0)]$ 
and $:\left[\partial_y\Phi(x=0)\right]^2:$.
At $g=1/2$ and $h\to 0$, both operators are irrelevant with
scaling dimension 2. A general combination of this sort can 
now be rewritten in terms of the 
underlying two Ising models.  By means of bosonization\cite{AKL}
we  obtain the equivalent combination 
$\lambda:\psi\partial_y\psi:+\mu:\xi\partial_y\xi:$.
By using the  same reasoning as in 
the previous paragraph, Eq.~(\ref{equivform}) yields
\begin{eqnarray}\label{newref}
P+Q&=&i\coth\left({\beta-\beta_B^3\over 2}-{i\pi\over 4}\right)\\ \nonumber
P-Q&=&-i\,\tanh\left({\beta-\beta_B^1\over 2}-{i\pi\over 4}\right)
\coth\left({\beta-\beta_B^2\over 2}-{i\pi\over 4}\right) \;.
\end{eqnarray}
In the limit of small bare couplings, 
$\exp(\beta_B^1)\propto h^2, \exp(\beta_B^2)\propto 1/\lambda,\exp(\beta_B^3)\propto 1/\mu$,
and therefore $\beta_B^1\ll \beta_B^2,\beta_B^3$. 
For temperatures $T\ll \exp(\beta_B^2),\exp(\beta_B^3)$, 
the $\coth$-factors are saturated, and we recover Eq.~(\ref{corresp}). For 
high temperatures, in contrast, one gets $P+Q\to i$, $P-Q\to -i$,
corresponding to Neumann (free) boundary conditions. 
The situation (\ref{newref}) is thus similar to the case 
$N=3$ for the Ising model. Notice however that in the special case
 $\mu=0$, we always get $P+Q=-i$, and the situation now 
looks like the case $N=2$ for the Ising model, with identical Dirichlet
(fixed) boundary conditions in the UV and IR.  The Neumann 
boundary condition is generally found only in the temperature regime
$\exp(\beta_B^1)\ll T \ll \exp(\beta_B^2)$. 

\begin{figure}
\epsfxsize=1.0\columnwidth
\epsffile{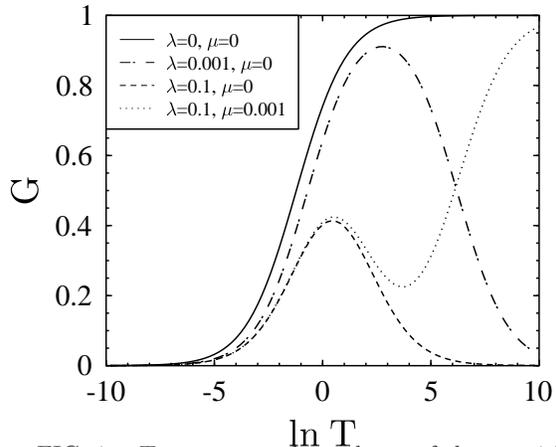}
\caption[]{\label{fig1} 
Temperature dependence of the $g=1/2$
conductance (\ref{conduc}) for $h=1$ and various $\lambda, \mu$.
}
\end{figure}

Following Ref.~\onlinecite{FLS}, the exact transport properties can  
be computed for arbitrary boundary interactions
once the reflection matrices are known.
The $g=1/2$ linear conductance in units of $g e^2/h$ is\cite{foot}
\begin{equation}\label{conduc}
G=\int_{-\infty}^\infty d\beta 
{1\over 1+e^{\beta}}{d\over d\beta}|Q(\beta+\ln T)|^2\;,
\end{equation}
where $Q$ is given by Eq.~(\ref{newref}).
This formula results from Ref.~\onlinecite{FLS}
after an integration by parts,
and holds provided $|Q|^2$ is not a constant, so there is a flow indeed.
Figure \ref{fig1} shows plots for Eq.~(\ref{conduc}).
In the absence of the irrelevant couplings, the conductance
goes from the perfect value $ge^2/h$ at high $T$
(Neumann boundary conditions) down to $G=0$ as $T\to 0$
(Dirichlet). 
With the irrelevant couplings, the situation is more complex.
For $\mu=0$, in the UV limit
Dirichlet boundary conditions ($G=0$) are approached again.
However, for finite $\mu$, the UV fixed point corresponds to 
Neumann boundary conditions, see Fig.~\ref{fig1}.

\begin{figure}
\epsfxsize=0.9\columnwidth
\epsffile{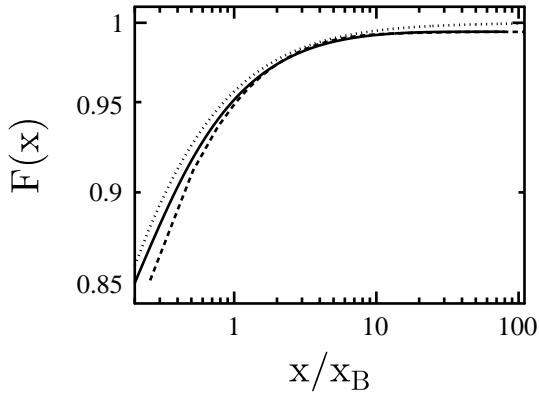}
\caption[]{\label{fig2} 
Monte Carlo results for the $g=1/2$ pinning function ($h=1$).
A hard energy cut-off $\omega_c=1/a$ was employed, and the
simulations were done at very low but finite temperature
$\omega_c/ T=1200$.  The exact $T=0$ result\cite{LLS} for $\lambda=0$
is given as dotted curve. 
Data for $\lambda=0$ are given as solid curve, and 
for $\lambda=0.6$ as dashed curve after 
rescaling $x_B$ by a factor 0.43. Statistical errors are
of the order of the thickness of the curves.
}
\end{figure}

Let us now discuss the effect of irrelevant operators in
a cut-off regularized theory.  We first consider the $g=1/2$ BSG model
(\ref{bdrsg}) and only add the  first of the above two operators, 
$\cos[2\sqrt{2\pi g} \Phi(x=0)]$, with a coupling constant $\lambda$.
Its effect is studied by Monte Carlo simulation for
the pinning function $F(x)$ of the Friedel oscillation.\cite{EG}
The $2k_F$-oscillatory charge
screening cloud $\rho(x)$ of a spinless Luttinger liquid around the impurity
at position $x=0$ is 
\begin{equation} \label{pinning}
\frac{\rho(x)-\rho_0}{\rho_0} = \cos(2k_F x+\eta_F) (a/x)^g F(x) \;,
\end{equation}
where $\rho_0=k_F/\pi$ and $\eta_F$ is a phase shift.
In the absence of the irrelevant coupling, $F$ depends on $x/x_B$ only,
where $x_B=a (h a)^{-1/(1-g)}$ is a lengthscale
corresponding to $\exp(\beta_B)$.
For $T=0$, energy scales are related to the
distance $x$ with $x\gg x_B$ meaning low energies. 
 In Eq.~(\ref{pinning}) we have introduced the lattice
spacing $a$ which serves as a high-energy cut-off.  
General arguments show\cite{LLS,EG} that $F(x)\to 1$ as $x\to \infty$.   
Figure~\ref{fig2} shows numerical results for the  $g=1/2$ 
pinning function.\cite{foot2}
From the $\lambda\neq 0$ data, one sees that in the low-energy
portion $x\gg x_B$, 
the additional irrelevant operator can simply be incorporated
by a rescaling of the lengthscale $x_B$ or the
associated rapidity $\beta_B$ (such a rescaling could also have been 
used in the fits of figure 1).  
A good estimate for the observed size of the rescaling can be 
obtained from a self-consistent harmonic approximation.\cite{EG}  
At high energy scales, $x\ll x_B$,
however, the curves do not scale on top of each other, 
in agreement with the general picture outlined above.

\begin{figure}
\epsfxsize=0.9\columnwidth
\epsffile{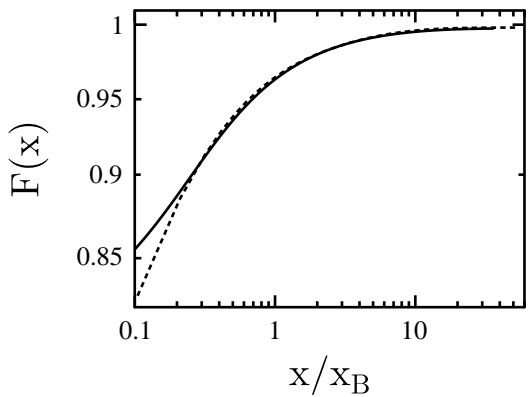}
\caption[]{\label{fig3} 
Monte Carlo results for the $g=1/5$ pinning function ($h=1$). 
Data for $\lambda=0$ are given as the solid curve, and 
for $\lambda=0.6$ as the dashed curve after
rescaling $x_B$ by a factor 0.19.
}
\end{figure}

The inclusion of additional irrelevant or less relevant 
terms is not restricted
to $g=1/2$.  First, recall that a  general consequence of integrability 
is the existence of a whole infinity of quantities in involution. 
A model perturbed by a combination of all these quantities
is then also integrable, and
the associated $R$ matrices should differ from the standard ones
only by the inclusion of CDD factors, giving rise to results 
similar to the Ising ones. As for adding operators that destroy integrability,
we have to resort to numerical simulations. We have studied  
the combination of the usual BSG term   $\cos\left[\sqrt{2\pi g}
\Phi(x=0)\right]$ and of 
$\cos[2\sqrt{2\pi g}\Phi(x=0)]$, which is not integrable away from $g=1/2$. 
 In Fig.~\ref{fig3}, we show data for
$g=1/5$, where the second operator  
is relevant (with scaling dimension 4/5), 
albeit of course  less relevant than the first.
We observe that at low energy scales ($x\gg x_B$),
 the coupling $\lambda$ has again no effect besides a
renormalization of $x_B$.
The scaling form of $F$ is altered only for $x\ll x_B$, giving rise to 
a picture that appears to be quite similar to the integrable one.

Our results should not lead to the belief that irrelevant 
operators are always of little importance.  
The integrable situation is quite special of course,
since it does not even require the introduction of a cut-off.
In non-integrable cases,  the necessary  cut-off might give rise to
more complex behaviors.
More importantly, the ``roaming'' RG trajectories we observed when approaching the UV fixed point  in the Ising model 
should serve as a warning that there are situations 
where irrelevant operators lead to counter-intuitive behavior,
sometimes even changing the nature of the  IR fixed point.
In the boundary sine-Gordon model, however,  this seems
never to be the case.

R.E.~and A.K.~acknowledge support by the Deutsche
Forschungsgemeinschaft (Bonn). 
H.S.~was supported by the DOE and the NSF (under the NYI program).


\begin{references}
\bibitem{foot0}
We have absorbed a parameter with dimension 
$L^{-1}$ into $\exp(\beta)$.
\bibitem{foot2} The minor discrepancy between our data and the exact  $T=0$
result is due to the finite temperature.
\end{references}
\end{document}